\documentclass[osajnl,twocolumn,showpacs,preprintnumbers]{revtex4}
\usepackage{epsfig}
\newcommand{\ave}[1]{\ensuremath{\langle #1 \rangle}}
\newcommand{\g}{\gamma}
\newcommand{\D}{\Delta}
\newcommand{\de}{\delta}
\begin{document}

\title{Entanglement in the above-threshold optical parametric oscillator}
\author{Alessandro S. Villar$^1$, Kati\'uscia N. Cassemiro$^1$, Kaled Dechoum$^2$, 
Antonio Z. Khoury$^2$, Marcelo Martinelli$^1$, and Paulo Nussenzveig$^1$}
\email{nussen@if.usp.br}
\address{$^1$ Instituto de F\'\i sica, Universidade de S\~ao Paulo,
Caixa Postal 66318, 05315-970 S\~ao Paulo, SP, Brazil \\ 
$^2$ Instituto de F\'\i sica da Universidade Federal Fluminense, Boa Viagem,  
24210-340, Niter\'oi, RJ, Brazil} 

\begin{abstract}
We investigate entanglement in the above-threshold Optical Parametric Oscillator, 
both theoretically and experimentally, and discuss its potential applications to 
quantum information. The fluctuations measured in the subtraction of signal 
and idler amplitude quadratures are $\Delta^2 \hat p_-=0.50(1)$, or 
$-3.01(9)$~dB, and in the sum of phase quadratures are $\Delta^2 \hat q_+=0.73(1)$, 
or $-1.37(6)$~dB. A detailed experimental study of the noise behavior as 
a function of pump power is presented, and discrepancies with theory are 
discussed. 
\end{abstract}

\maketitle

\section{Introduction}
\label{sec:intro} 

The Optical Parametric Oscillator (OPO) has been studied since the 
1960's\cite{kingston62,grahamhaken68}. Already in the 1980's it was 
recognized as an important tool in quantum optics, for the generation 
of squeezed states of light~\cite{kimble,heidmann}. It was also 
recognized as a suitable system for the demonstration of continuous variable 
(CV) entanglement, by Reid and Drummond, in 1988~\cite{reiddrummprl88}, 
where above-threshold operation was considered. In the early 1990's, 
CV entanglement was indeed demonstrated for the first time in an OPO, 
although operating below threshold~\cite{kimbleepr}. The OPO has since 
been used in several applications in CV quantum 
information~\cite{vanloockbraunrmp05,bowen03,schori02,kimbleteleport,entangswap04}. 
Entanglement in the above-threshold OPO, on the other hand, remained an 
experimental challenge until 2005, when it was first observed by Villar 
{\it et al.}~\cite{prlentangtwinopo}, and subsequently by two other 
groups~\cite{optlettpeng,pfisterentang}. 

Bipartite continuous variable entanglement can be demonstrated by a violation of 
the following inequality, obtained independently by Duan {\it et al.}~\cite{dgcz} 
and Simon~\cite{simon}: 
\begin{equation}
\Delta^2 \hat p_-+\Delta^2 \hat q_+\geq 2 \,,
\label{dgczcrit}
\end{equation}
where $\hat p_-=(\hat p_1-\hat p_2)/\sqrt{2}$ and 
$\hat q_+=(\hat q_1+\hat q_2)/\sqrt{2}$ are EPR-like 
operators constructed by combining operators of each subsystem. We choose 
$\hat p_j$ and $\hat q_j$, $j\in\{0,1,2\}$, as the amplitude and phase 
quadrature operators of the pump, signal and idler fields, respectively, 
which obey the commutation relations $[\hat p_j,\hat q_k]=2 i \delta_{jk}$. 
Any separable system must satisfy Eq.~(\ref{dgczcrit}): violation is an 
unequivocal signature of entanglement.

Entanglement between the intense signal and idler beams generated by an 
above-threshold OPO can be physically understood as a consequence of 
energy conservation in the parametric process. On one hand, pump photons are 
converted into pairs of signal and idler photons, leading to strong 
intensity correlations; on the other hand, the sum of frequencies of 
signal and idler photons is fixed to the value of pump frequency, 
leading to phase anti-correlations. The difficulty of measuring phase fluctuations 
was largely responsible for the long time between the prediction and the first 
observation of entanglement in the above-threshold OPO. The technique 
we used to measure phase fluctuations consists of reflecting each field 
off an empty optical cavity, as explained in Ref~\cite{optcomm04}. 

The value of Eq.~(\ref{dgczcrit}) obtained in the first 
demonstration of entanglement was 1.41(2), with squeezing 
observed in both EPR-like operators, $\Delta^2\hat p_-=0.59(1)$ and 
$\Delta^2\hat q_+=0.82(2)$~\cite{prlentangtwinopo}. Nevertheless, such a result 
could only be achieved very close to threshold, otherwise the phase sum 
$\Delta^2\hat q_+$ would present excess noise, increasing 
with pump power relative to threshold $\sigma=P_0/P_{\mbox{\scriptsize th}}$. 
This strange behavior, also observed by other groups~\cite{claude10db}, is not 
predicted by the standard linearized OPO theory, for a shot noise limited pump beam. 
According to this model, entanglement should exist for all values of $\sigma$, 
although the degree of entanglement should decrease for increasing $\sigma$. 
This presented an additional complication for the first demonstration of 
entanglement in the above-threshold OPO. 

In this paper, we present new improved results of entanglement in the 
above-threshold OPO, together with a theoretical and experimental study 
of this unexpected excess phase sum noise. The paper is organized as follows. 
We begin by describing the linearized model for the OPO and its predictions 
for a shot noise limited pump beam. This model includes losses and also allows 
for nonvanishing detunings of pump, signal, and idler modes with respect to 
the OPO cavity. We then present a full-quantum treatment, neglecting losses and 
for zero detunings. Even after eliminating the linearization approximation, the 
theory does not predict the observed excess noise. The experiment is described next, 
and we present measurements of sum and difference of quadratures' fluctuations, 
as a function of $\sigma$. The excess noise in the phase sum can be related 
to pump noise generated inside the OPO cavity, as we will see. We finally present 
our currently best measurement of two-color squeezed-state entanglement. We 
conclude by mentioning applications of this entanglement in quantum information. 

\section{Theoretical description of the OPO} 
\label{sec:theory}

The optical parametric oscillator consists of three modes of the electromagnetic 
field coupled by a nonlinear crystal, which is held inside an optical cavity. The 
OPO is driven by an incident pump field at frequency $\omega_{0}$. Following the usual 
terminology, the downconverted fields are called signal and idler, of frequencies 
$\omega_{1}$ and $\omega_{2}$, where, by energy conservation, $\omega_{0}=\omega_{1}
+\omega_{2}$. We will treat here the case of a cavity which is triply resonant for 
$\omega_0$, $\omega_{1}$, and $\omega_{2}$. Each field is damped via the cavity 
output mirror, thereby interacting with reservoir fields. The effective second-order 
nonlinearity of the crystal is represented by the constant $\chi$. 

Reid and Drummond investigated the correlations in the nondegenerate OPO (NOPO) 
both above~\cite{MDacima} and below threshold~\cite{PDbaixo}. 
In the above threshold case, they studied the effects of phase 
diffusion in the signal and idler modes, beginning with the 
positive P-representation equations of motion for the interacting 
fields~\cite{+P1,+P2}. Changing to intensity and phase variables, they 
were able to show that output quadratures could be chosen which 
exhibited fluctuations below the coherent state level and also 
Einstein-Podolsky-Rosen (EPR) type correlations. In the below 
threshold case, a standard linearized calculation was sufficient to 
obtain similar correlations. In the limit of a rapidly decaying pump 
mode, Kheruntsyan and Petrosyan were able to calculate exactly the 
steady-state Wigner function for the NOPO, showing clearly the 
threshold behavior and the phase diffusion above this level of 
pumping~\cite{barbudos}. 

We begin by describing the linearized model, and then proceed to 
calculate noise spectra beyond linearization.
\subsection{The linearized model}

The equations describing the evolution of signal, idler, and pump amplitudes, 
$\alpha_j$, inside the triply resonant OPO cavity are given below~\cite{optcomm04}. 
They are obtained by writing the density operator equation of motion in the Wigner 
representation, and then searching for a set of equivalent Langevin equations.
\begin{widetext}
\begin{eqnarray}
\tau\frac{d}{dt} \alpha_0 & = & -\g'_0(1-i\D_0)\,\alpha_0-2\chi^*\alpha_1\alpha_2+ 
\sqrt{2\g_0}\,\alpha_0^{in}+\sqrt{2\mu_0}\,\de v_0  \nonumber \\ 
\tau\frac{d}{dt} \alpha_1 & = & -\g'(1-i\D)\,\alpha_1+2\chi\,\alpha_0\alpha_2^*+ 
\sqrt{2\g}\,\de u_1+\sqrt{2\mu}\,\de v_1 \label{eqopo} \\ 
\tau\frac{d}{dt} \alpha_2 & = & -\g'(1-i\D)\,\alpha_2+2\chi\,\alpha_0\alpha_1^*+ 
\sqrt{2\g}\,\de u_2+\sqrt{2\mu}\,\de v_2 \;, \nonumber 
\end{eqnarray}
\end{widetext}
where $\g$ and $\g_0$ are half the transmissions of the mirrors, 
$\g'$ and $\g'_0$ are the total intracavity losses, 
$\mu=\g'-\g$ and $\mu_0=\g'_0-\g_0$ are the spurious intracavity losses, 
$\D$ and $\D_0$ are the detunings of the OPO cavity relative to the 
fields' central frequencies, and $\tau$ is the cavity roundtrip time. 
We have considered here that $\g_1=\g_2=\g$ and  $\g'_1=\g'_2=\g'$. The 
parameter $\chi$ is the effective second-order nonlinearity. The terms 
$\de u_j$ and $\de v_j$ are vacuum fluctuations associated to the losses 
from the mirrors' transmissions and from spurious sources, respectively. 
In the case of the intracavity pump mode, the fluctuations that come from 
the mirror transmission are due to the quantum fluctuations of the input 
pump laser beam, $\de\alpha_0^{in}=\de p_0^{in}+i\,\de q_0^{in}$.

Linearization consists in writing $\alpha_j(t)= e^{i\phi_j}(p_j+\de p_j(t)+ 
i\de q_j(t))$ and ignoring terms that involve products of fluctuations in 
the equations. Here $\ave{\alpha_j}=p_j e^{i\phi_j}$ is each field's mean 
amplitude, with $p_1=p_2\equiv p$ for equal overall intracavity losses in 
signal and idler, $\de p_j(t)$ is the amplitude fluctuation, and $\de q_j(t)$ 
is the phase fluctuation. Taking the average of the resulting equations gives 
information on the mean values of the fields. We may then separate the 
fluctuating part in real and imaginary contributions in order to obtain the 
equations of evolution for the quadratures of the fields. Defining 
$\de q_\pm=(\de q_1\pm\de q_2)/\sqrt{2}$ and $\de p_\pm=(\de p_1\pm\de p_2)/\sqrt{2}$ as 
the normalized sum/subtraction of signal and idler amplitude and phase quadratures, 
we write the above equations in terms of the EPR variables:
\begin{widetext}
\begin{eqnarray}
\tau\frac{d}{dt} \,\de p_- & = & -2\g'\,\de p_- + 
   \sqrt{2\g}\,\de u_{p-} + \sqrt{2\mu}\,\de v_{p-} \nonumber \\ 
\tau\frac{d}{dt} \,\de q_- & = & 2\D\g'\,\de p_- + 
   \sqrt{2\g}\,\de u_{q-} + \sqrt{2\mu}\,\de v_{q-} \nonumber \\ 
\tau\frac{d}{dt} \,\de p_+ & = & -2\D\g'\,\de q_+ + 
   \sqrt{2}\g'\beta\,\de p_0 + \sqrt{2}\D\g'\beta\,\de q_0 + 
   \sqrt{2\g}\,\de u_{p+} + \sqrt{2\mu}\,\de v_{p+} \label{eqopolinear} \\
\tau\frac{d}{dt} \,\de q_+ & = & -2\g'\,\de q_+ - 
   \sqrt{2}\D\g'\beta\,\de p_0 + \sqrt{2}\g'\beta\,\de q_0 + 
   \sqrt{2\g}\,\de u_{q+} + \sqrt{2\mu}\,\de v_{q+} \nonumber \\ 
\tau\frac{d}{dt} \,\de p_0 & = & -\sqrt{2}\g'\beta\,\de p_+ +
   \sqrt{2}\D\g'\beta\,\de q_+ - \g'_0\,\de p_0 - \D_0\g'_0\,\de q_0 + 
   \sqrt{2\g_0}\,\de p_0^{in} + \sqrt{2\mu_0}\,\de v_{p0} \nonumber \\ 
\tau\frac{d}{dt} \,\de q_0 & = & -\sqrt{2}\D\g'\beta\,\de p_+ - 
   \sqrt{2}\g'\beta\,\de q_+ + \D_0\g'_0\,\de p_0 - \g'_0\,\de q_0 + 
   \sqrt{2\g_0}\,\de q_0^{in} + \sqrt{2\mu_0}\,\de v_{q0} \;, \nonumber
\end{eqnarray}
\end{widetext}
where $\beta=p/p_0$ is the ratio between the intracavity amplitudes of 
downconverted and pump fields. Noise spectra of the transmitted fields 
are calculated by solving the above equations in Fourier space. We 
define $S_{p\pm}$ and $S_{q\pm}$ as the noise spectra of the operators 
$\hat p_\pm$ and $\hat q_\pm$, respectively. 

It is clear from Eq.~(\ref{eqopolinear}) that the subtraction of 
quadratures' subspace decouples from the others, so that $S_{q-}$ and 
$S_{p-}$ depend only on the ratio of losses through the output cavity 
mirror to the total intracavity losses, and on the analysis frequency 
$\Omega$. These fluctuations do not depend on pump power, and are in 
a minimum uncertainty state, $S_{p-} \times S_{q-}=1$, if $\g=\g'$.

\begin{figure}[ht]
\centering
\epsfig{file=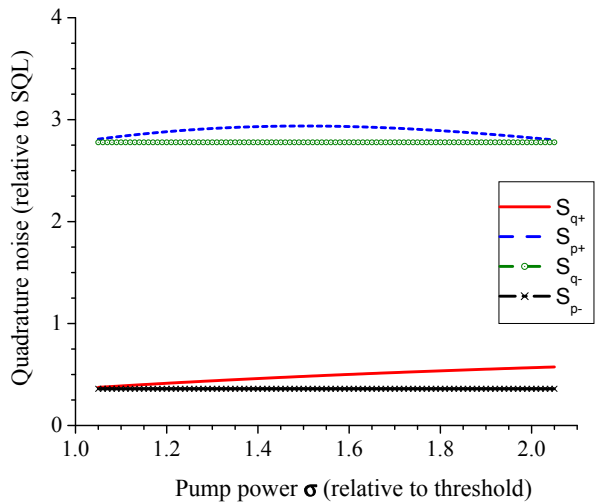,scale=0.35}
\caption{Prediction of the linearized theory for fluctuations in the sum/subtraction 
of field quadratures as a function of $\sigma$ for a shot noise limited pump beam. 
Full line: $S_{q+}$; dashed line: $S_{p+}$; line + crosses: $S_{p-}$; line + 
circles: $S_{q-}$}
\label{ruidoxdcteor}
\end{figure}

On the other hand, the sum of quadratures and pump fields' subspaces are 
connected. This directly implies that excess noise in the pump beam degrades
signal-idler entanglement, and can even destroy it~\cite{optcomm04}. The behavior 
of the twin beams' fluctuations as functions of pump power relative to threshold $\sigma$, 
for a shot noise limited pump, is presented in Fig.~\ref{ruidoxdcteor}. The maximum 
squeezing of $S_{q+}$ occurs at threshold, and approaches shot noise for higher 
pump powers. 

These behaviors change in the presence of excess noise in the pump. 
In this case, both $S_{q+}$ and $S_{p+}$ increase from their values at 
threshold. In particular, $S_{q+}$ goes from squeezing to excess noise.
The point where it crosses the shot noise value solely depends on the
amount of excess phase noise present in the pump beam. For this reason, it 
was necessary to filter the pump field in the experiment, in order to observe 
entanglement.

\subsection{Noise spectra beyond the linearized model}
We present here a comparison between the linearized approach 
to the quantum noise in the OPO and the numerical integration of the 
quantum stochastic equations in the positive P-representation. This will 
help us to eliminate the linearization procedure as the reason for the 
discrepancy between the theoretical prediction of squeezing and the 
experimentally observed excess phase noise for $\sigma > 1.2$. We shall 
follow the procedure used in Ref.~\cite{nos} 
Although exact Heisenberg equations of motion can be found for this 
system, it is, at the very least, extremely difficult to solve nonlinear 
operator equations. We therefore develop stochastic equations of motion 
in the positive P-representation, which in principle give access to any 
normally-ordered operator expectation values we may wish to calculate.
To find the appropriate equations, we proceed via the master and Fokker-Planck 
equations. Using the standard techniques for elimination of the baths~\cite{hjc}, 
we find the zero-temperature master equation for the reduced density operator.
The master equation may be mapped onto a Fokker-Planck equation~\cite{Crispin} 
for the positive-P pseudoprobability distribution.  
The cavity damping rates at each frequency are $\gamma^D_{j}=2\g_j/\tau$, 
with $\gamma_{1}=\gamma_{2}= \gamma$. We further define $\gamma_{r}=\gamma_{0}/\gamma$. 
In order to apply perturbation theory, we introduce a normalized coupling constant, 
\begin{equation}
g = \frac{\chi}{\gamma^D \sqrt{2\gamma_{r}}}\;,
\label{eq:a17}
\end{equation}
which will be a power expansion parameter. Moreover, it will be useful to work 
with the scaled quadratures
\begin{eqnarray}
&&x_{0} = g \sqrt{2 \gamma_{r}}\,p_{0} \quad , \qquad q_{0} = g \sqrt{2 \gamma_{r}}
\,q_{0}, \nonumber \\
&&x_{+} = g\,p_{+} \quad , \; \qquad \qquad y_{+} = g\,q_{+}, \nonumber \\
&&x_{-} = g\,p_{-} \quad , \; \qquad \qquad y_{-} = g\,q_{-} \quad ,
\label{eq:a16}
\end{eqnarray}
in order to render the stochastic equations amenable to perturbation.
The stochastic equations for the scaled EPR variables become
\begin{widetext}
\begin{eqnarray}
\frac{d x_{0}}{d T} &=& -\gamma_{r} \left[ x_{0} - 2\sqrt{\sigma} + 
   \frac{1}{2} \left( x_{+}^{2} -  x_{-}^{2} - y_{+}^{2} + y_{-}^{2} \right) 
   \right] \;, \nonumber \\ 
\frac{d y_{0}}{d T} &=& -\gamma_{r} \left[ y_{0} + x_{+} y_{+} - 
   x_{-}y_{-} \right] \;, \nonumber \\ 
\frac{d x_{-}}{d T} &=&  - x_{-} -  \frac{1}{2} \left[ x_{0} x_{-} 
   + y_{0}y_{-} \right] + \frac{g}{\sqrt{2}} \left[ \sqrt{x_{0} + 
   i y_{0}}\;\xi_{-} + \sqrt{x_{0} - i y_{0}}\; \xi_{-}^{+} \right] \;, \label{eq:a18} \\
\frac{d y_{+}}{d T} &=&  - y_{+} +  \frac{1}{2} \left[ y_{0} x_{+} - 
   x_{0}y_{+} \right] - i \frac{g}{\sqrt{2}} \left[ \sqrt{x_{0} + 
   i y_{0}}\;\xi_{+} - \sqrt{x_{0} - i y_{0}}\; \xi_{+}^{+}\right] \;,\nonumber \\
\frac{d x_{+}}{d T} &=&  - x_{+} + \frac{1}{2} \left[ x_{0} x_{+} + 
   y_{0}y_{+} \right] + \frac{g}{\sqrt{2}} \left[ \sqrt{x_{0} + 
   i y_{0}}\; \xi_{+} + \sqrt{x_{0} - i y_{0}}\; \xi_{+}^{+}\right] \;, \nonumber \\
\frac{d y_{-}}{d T} &=& - y_{-} + \frac{1}{2} \left[ x_{0} y_{-} 
   - y_{0}x_{-} \right] - i \frac{g}{\sqrt{2}} \left[ \sqrt{x_{0} + 
   i y_{0}}\;\xi_{-} - \sqrt{x_{0} - i y_{0}}\; \xi_{-}^{+} \right] \nonumber\;,
\end{eqnarray}
\end{widetext}
where $T=\gamma^D\,t$ is time in units of the cavity lifetime for the 
down-converted fields. The functions $\xi_{\pm}(T)$ and $\xi_{\pm}^+
(T)$ are independent Langevin forces with the following nonvanishing 
correlation functions:
\begin{eqnarray}
\langle \xi_{+}(T) \xi_{+}(T^\prime) \rangle &=&\langle 
\xi_{+}^{+}(T) \xi_{+}^{+}(T^\prime) \rangle = \delta 
(T - T^\prime)\;, \nonumber\\
\langle  \xi_{-}(T) \xi_{-}(T^\prime) \rangle &=&\langle 
\xi_{-}^{+}(T) \xi_{-}^{+}(T^\prime T) \rangle = 
-\delta (T - T^\prime)\;.
\end{eqnarray}
We notice the symmetry properties of the stochastic equations~(\ref{eq:a18}). 
In fact, it is easy to verify that the equations of motion 
are unchanged by the transformation $x_-\leftrightarrow y_+\,$ 
and $x_+\leftrightarrow -\,y_-\,$. Of course, all noise terms 
appearing in Eqs.~(\ref{eq:a18}) are statistically equivalent. 
Therefore, these equations should not change the symmetries of the 
initial values chosen for $x_+$ and $y_-$.
In order to provide a comparison between the linearized model 
and the full stochastic integration, we will use a perturbation 
expansion of the positive P-representation of the dynamical 
equations. This allows us to include quantum effects in a systematic 
fashion~\cite{ndturco}. We first introduce a formal 
perturbation expansion in powers of the parameter $g$,
\begin{eqnarray}
&& x_{k} = \sum_{n=0}^{\infty} g^{n} x_{k}^{(n)}, \nonumber \\
&& y_{k} = \sum_{n=0}^{\infty} g^{n} y_{k}^{(n)}.
\label{eq:a20}
\end{eqnarray}
The series expansion written in this way has the property that the 
zeroth order term corresponds to the classical field of order $1$ in the unscaled 
quadrature, while the first order term is related to quantum 
fluctuations of order $g$, and the higher order terms correspond to nonlinear 
corrections to the quantum fluctuations of order $g^{2}$ and greater. The 
stochastic equations are then solved by the technique of matching powers of 
$g$ in the corresponding time evolution equations. 
The steady state solutions $x_{js}$ of the zeroth order give the operation 
point of the OPO and describe its macroscopic behavior. For triply resonant 
operation, the expressions for the steady state are quite simple:
\begin{eqnarray}
x_{0s}&=&2\;,\nonumber\\
x_{+s}&=&2\,\left(\sqrt{\sigma}-1\right)^{1/2}\;,\nonumber\\
x_{-s}&=&0\;,\label{sstate} \\
y_{0s}&=&y_{+s}=y_{-s}=0\;.\nonumber
\end{eqnarray}
The first order equations are often used to predict squeezing in a 
linearized fluctuation analysis. They are non-classical in the sense that 
they can describe states without a positive-definite Glauber-Sudarshan 
P-distribution~\cite{Roy,Sudarshan}, but correspond to a simple form of linear 
fluctuation which has a Gaussian quasi-probability distribution. A full 
quantum description of the OPO dynamics can be obtained by numerical 
integration of the stochastic equations~(\ref{eq:a18}), and can be compared 
to analytical expressions obtained from the linearized approach. Taking the 
first order terms and using the steady state solutions given by Eqs.~(\ref{sstate}), 
we can write the following equations for the linear quantum fluctuations,
\begin{eqnarray}
\frac{d x_{0}^{(1)}}{d T} &=& -\gamma_{r} \left[ x_{0}^{(1)} + 
   2\,\left(\sqrt{\sigma}-1\right)^{1/2}x_{+}^{(1)} \right] \;, \nonumber \\ 
\frac{d y_{0}^{(1)}}{d T} &=& -\gamma_{r} \left[ y_{0}^{(1)} +
   2\,\left(\sqrt{\sigma}-1\right)^{1/2} y_{+}^{(1)} \right] \;, \nonumber \\ 
\frac{dx_{+}^{(1)}}{d T} &=& - \left(\sqrt{\sigma}-1\right)^{1/2}\,x_{0}^{(1)} 
   +\left( \xi_{+} + \xi_{+}^{+} \right) \;, \label{eq:firstordernoise} \\
\frac{dx_{-}^{(1)}}{d T} &=& - 2\,x_{-}^{(1)} + \left( \xi_{-} + \xi_{-}^{+} 
   \right) \;, \nonumber \\
\frac{dy_{+}^{(1)}}{d T} &=& - 2\,y_{+}^{(1)} + \left(\sqrt{\sigma}-1\right)^{1/2}
   y_0^{(1)}-i \left( \xi_{+} - \xi_{+}^{+} \right) \;, \nonumber \\
\frac{dy_{-}^{(1)}}{d T} &=& -i \left( \xi_{-} - \xi_{-}^{+} \right) \nonumber\;. 
\end{eqnarray}
The linear coupled stochastic equations obtained agree with Eqs.~(\ref{eqopolinear}), 
for zero detunings and no spurious losses. From them, we may readily calculate the 
steady state averages of the first-order corrections and use that to compute the 
linearized fluctuations. Notice that under the linear approximation $y_-$ 
becomes a purely diffusive variable (phase diffusion). 
In an experimental situation, the noise spectra outside the cavity are 
generally the quantities of interest. We will therefore proceed to 
analyze the problem in frequency space, via Fourier decomposition of 
the fields. The first order stochastic equations may be rewritten in the frequency 
domain so that we may calculate the spectra of the squeezed and anti-squeezed 
field quadratures. 
The solutions for the noise of the squeezed operators, $\hat p_{-}$ and $\hat q_{+}$, are:
\begin{equation}
S_{p-}(\Omega^\prime) = 1 - \frac{1}{\Omega^{\prime 2} + 1}
\label{eq:sp-}
\end{equation}
and
\begin{widetext}
\begin{equation}
S_{q+}(\Omega^\prime) = 1 - \frac{(4\,\Omega^{\prime 2}+\gamma_r^2)^2} 
{\Omega^{\prime 2}\left[ 4\,\Omega^{\prime 2} + \gamma_r^2 - 2\,\gamma_r\, 
\left(\sqrt{\sigma}-1\right)\right]^2 + \left[ 4\,\Omega^{\prime 2} + 
\gamma_r^2\,\sqrt{\sigma}\right]^2}\;,
\label{eq:sq+}
\end{equation}
\end{widetext}
where $\Omega^\prime=\Omega/\g^D$ is the analysis frequency in units of 
the cavity bandwidth.
Under the limits of the linearized approach, the results of the 
noise spectra are independent of the phase space representation 
employed. Therefore, these results coincide with the usual ones 
obtained with the Wigner representation.
The spectra given by Eqs.~(\ref{eq:sp-}) and (\ref{eq:sq+}) can now be compared 
with those found via stochastic integration of the full equations of motion~(\ref{eq:a18}) 
in the positive P-representation. The nonlinear spectra are calculated by Fourier transform 
of the stochastic integration, which must be performed numerically. A somewhat subtle 
point arises here: the nonlinear Eqs.~(\ref{eq:a18}) have more than one possible 
steady-state solution. Thus, for a fair comparison with the linearized spectra, 
it is necessary to choose the same steady-state. By doing this, we verified that 
both predictions, in the above-threshold OPO, agree within a good numerical 
precision. Therefore, we conclude that possible limitations of the linearized model 
for dealing with the OPO dynamics under phase diffusion do not account for the 
experimentally observed excess noise of $\hat q_+$. 
\section{Experiment}
\label{sec:exp}

\begin{figure}[ht]
\centering
\epsfig{file=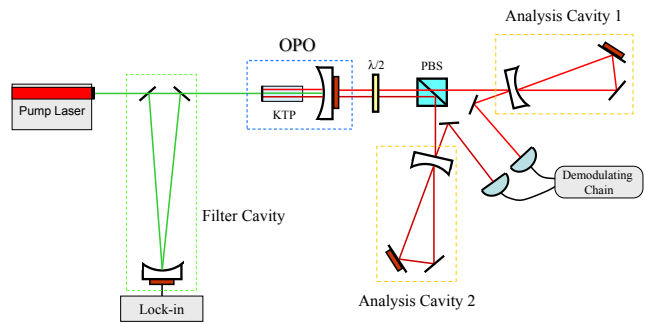,scale=0.35}
\caption{Sketch of the experimental setup.}
\label{setup}
\end{figure}

\begin{figure}[ht]
\centering
\epsfig{file=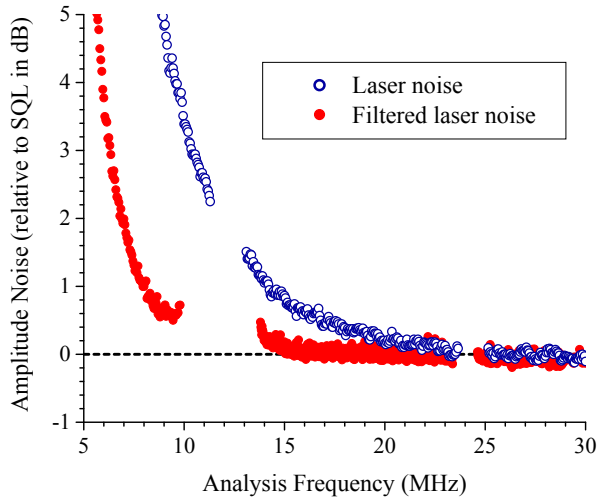,scale=0.35}
\caption{Measurement of the pump noise, as a function of the analysis 
frequency. Open circles: unfiltered laser noise; full circles: laser 
noise at the output of the filter cavity. In view of the large excess noise 
at 12~MHz and its second harmonic, we suppressed those frequencies from 
the data.}
\label{pumpnoise}
\end{figure}

Our system is a triply resonant type-II OPO operating above threshold. The 
experimental setup is depicted in Fig.~\ref{setup}. The pump beam is 
a diode-pumped doubled Nd:YAG laser (Innolight Diabolo) with 900~mW output 
power at 532~nm. A secondary output at 1064~nm is used for alignment purposes. 
Since  the pump beam presents excess noise for frequencies as high as 20~MHz, 
a filter cavity is necessary. Our filter cavity has a bandwidth of 2.4~MHz 
and assures that the pump laser is shot noise limited for analysis frequencies 
higher than 15~MHz (see Fig.~\ref{pumpnoise}). We measured the laser phase 
noise by reflecting the beam off an empty cavity, in the same way we measure 
phase noise of the downconverted beams. The phase noise equals the intensity 
noise, except at a frequency of 12~MHz, where there is very big phase noise, 
owing to a frequency modulation inside the Diabolo laser, for stabilization 
purposes. This excess noise saturates our electronics and prevents measurements 
for analysis frequencies close to 12~MHz and also to its second harmonic, 
24~MHz, as can be seen in Fig.~~\ref{pumpnoise}. The OPO cavity is a linear 
semi-monolithic cavity composed of a flat input mirror, directly deposited 
on one face of the nonlinear crystal, with 93\% reflectivity at 532~nm and 
high reflectivity ($>99.8$\%) at 1064~nm, and a spherical output mirror 
(50~mm curvature radius) with high reflectivity at 532~nm ($>99.8$\%) and 
96\% reflectivity at 1064~nm. The nonlinear crystal is a 10~mm-long Potassium 
Titanyl Phosphate (KTP) from Litton. Threshold power is 12~mW. 

Signal and idler beams are separated by a polarizing beam splitter (PBS) and 
sent to detection, which consists of a ring cavity and a photodetector (Epitaxx 
ETX 300) for each beam. Overall detection efficiency is $\eta=80(2)\%$. Both 
analysis cavities have bandwidths of 14~MHz, allowing for a complete conversion 
of phase to amplitude noise for analysis frequencies higher than 20~MHz. 
Measurements are taken at analysis frequency equal to 27~MHz. In order to 
access the same quadrature for both beams, the two cavities must be detuned by 
the same amount at the same time. By scanning the detunings synchronously, we 
can measure all quadratures of the twin beams. In particular, we can easily 
select the amplitude (off resonance) or phase (detuning equal to half the 
bandwidth) quadratures\cite{galatola}. 

Data acquisition is carried out by a demodulating chain, which mixes the 
photocurrents from each detector with a sinusoidal electronic reference 
at the analysis frequency and filters the resulting low frequency signal. 
The demodulated photocurrent fluctuations are sampled at 600~kHz repetition rate 
by an A/D card connected to a personal computer. The variances of these 
fluctuations are then computed taking groups of 1000 points, resulting in 
something proportional to the photocurrents' power spectrum at the analysis 
frequency. At the end, measured variances are normalized to the shot noise 
standard quantum level (SQL).

\subsection{Fluctuations as a function of $\sigma$}

The input pump field is guaranteed to be shot noise limited for frequencies above 15~MHz 
after being transmitted through the filter cavity. Even before being filtered, 
pump field is shot noise limited above 25~MHz, as shown in Fig.~\ref{pumpnoise}. 
Nevertheless, we observed excess noise in the sum of phases of signal and idler beams, 
preventing the violation of the inequality given in Eq.~\ref{dgczcrit}, except for 
pump powers very close to threshold\cite{prlentangtwinopo}. 

\begin{figure}[ht!]
\centering
\epsfig{file=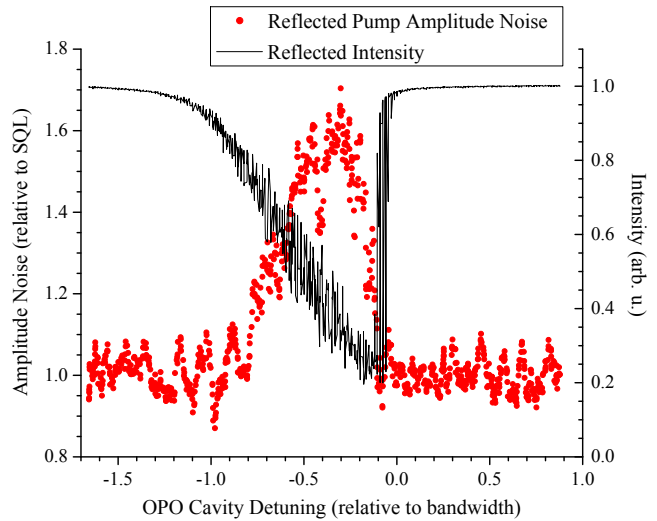,scale=0.35}
\caption{Intensity noise of the reflected pump beam, as a function of the 
detuning of the OPO cavity. The excess noise observed is peaked for $\D_0$ 
close to half the OPO cavity bandwidth. The asymmetry in the mean field signal 
is due to thermal bistability. The analysis frequency is 27~MHz. Circles: reflected 
pump noise; full line: reflected average intensity}
\label{pumpnoiseref}
\end{figure}

As seen in section~\ref{sec:theory}, from the theoretical description of 
the OPO, excess noise in the pump beam would generate excess noise in the 
phase sum of the twin beams. Yet, how could that be the case if we 
carefully measured the input pump to be shot noise limited? By following 
this single lead, it is natural to examine the noise properties of the 
pump beam reflected from the OPO cavity. This was done by scanning the 
OPO cavity, for crystal temperatures such that there was no parametric 
oscillation (triple resonance depends sharply on crystal temperature 
and can be easily avoided). Since the incident beam is shot noise limited, 
could there be excess noise generated inside the cavity containing the 
KTP crystal? We did indeed find excess noise in the reflected pump's 
amplitude (Fig.~\ref{pumpnoiseref}) and phase quadratures. The maximum 
values, for $\sigma=1$, were $S^{\mathrm{R}}_{p0}= 1.8(1)$ and 
$S^{\mathrm{R}}_{q0}=4.5(3)$. 

At present, we can still not claim to fully understand the origin 
of this excess noise. We verified, of course, that no such noise is 
generated in an empty cavity (which would also invalidate the measurements 
we perform with the analysis cavities for the twin beams). We also checked 
whether this effect depended on $\chi$ and would thus be directly related to 
the parametric process. For a polarization of the incident beam orthogonal 
to the usual polarization, phase matching can not be fulfilled, 
and no downconversion can occur. The noise in the reflected beam did not 
show any significant dependence on the incident polarization. It does, however,  
increase for increasing power of the incident beam. We can speculate that 
this can be a result of photon absorption by the crystal at 532~nm (which 
is at the origin of the thermal bistability observed in Fig.~\ref{pumpnoiseref}), 
with subsequent relaxation by spontaneous emission or non-radiative processes. 
This may give rise to an intensity-dependent refractive index, yielding 
phase and amplitude modulation at 532~nm. We are currently investigating these 
possibilities. 

\begin{figure}[htbp!]
\centering
\epsfig{file=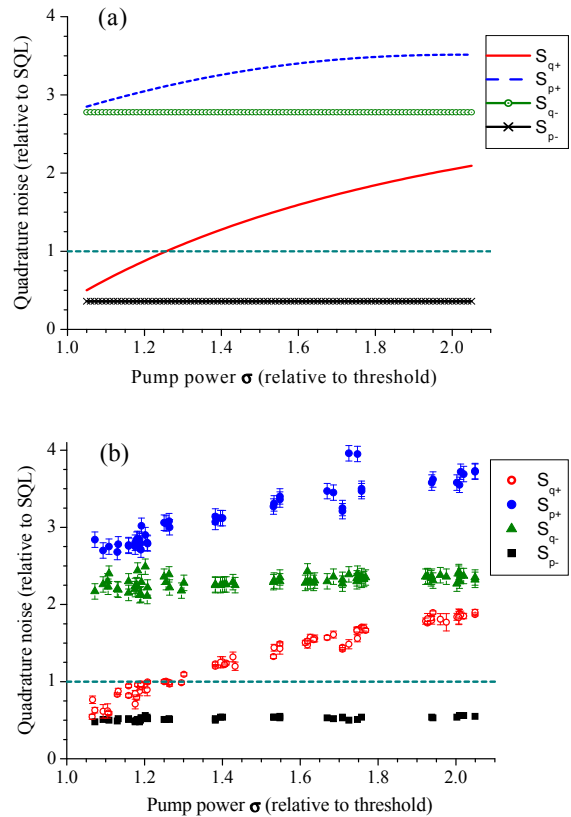,scale=0.45}
\caption{Noise behavior as a function of $\sigma$. In part (a), we 
present the predictions of the linearized model, for an input pump beam 
with $S_{p0}=1.5$ and $S_{q0}=5.5$; dashed line: $S_{p+}$; full line + open 
circles: $S_{q-}$; full line: $S_{q+}$; full line + crosses: $S_{p-}$; SQL = 1.0 
is indicated by a dashed line. In part (b), experimental results are shown for 
$\sigma$ ranging from 1.06 to 2.2. full circles: $S_{p+}$; triangles: $S_{q-}$; 
open circles: $S_{q+}$; squares: $S_{p-}$; SQL = 1.0 is indicated by a dashed line.}
\label{ruidoxdcexp}
\end{figure}

As a first approximation, in order to see whether this would account for 
the behavior of $\Delta^2 \hat p_-$, $\Delta^2 \hat q_-$, $\Delta^2 \hat p_+$, 
and $\Delta^2 \hat q_+$, as a function of $\sigma$, we simply 
added excess noise to the input pump beam in the linearized OPO theory. 
In Fig.~\ref{ruidoxdcexp}, we compare the results from the model, with 
incident $S_{p0}=1.5$ and $S_{q0}=5.5$, to the measured data. 
Signal and idler powers varied from 0.4mW up to 5.5mW each during the 
experiment, corresponding to pump powers between 13mW and 26mW, 
or $1.06<\sigma<2.2$. As expected, noises corresponding to the subtraction 
subspace, $\Delta^2 \hat p_-$ and $\Delta^2 \hat q_-$, are independent of pump 
power. But $\Delta^2 \hat q_+$ is very sensitive to $\sigma$, as well 
as $\Delta^2 \hat p_+$ to a smaller degree. The agreement with the theoretical 
model is surprisingly good. This is a strong indication that the intracavity 
pump excess noise is the main responsible for the excess noise in $\Delta^2 \hat q_+$.

\subsection{Two-color entanglement}

The sum of phases' noise is squeezed very close to threshold, and
squeezing is degraded with increasing pump power. $\Delta^2 \hat q_+$
crosses the shot noise level approximately at $\sigma=1.20$, from squeezing to
anti-squeezing, although only below $\sigma=1.15$ can squeezing be 
observed with certainty. 

\begin{figure}[ht!]
\centering
\epsfig{file=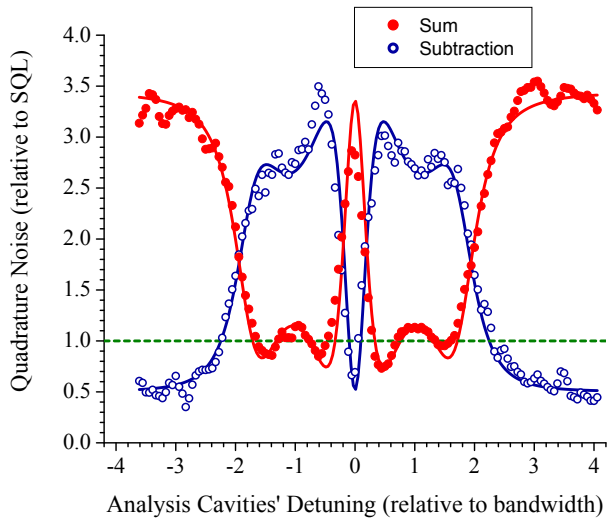,scale=0.35}
\caption{Sum (full circles) and difference (open circles) of quadratures' noise, 
measured as a function of the analysis cavities' detuning. Squeezed-state 
entanglement can be directly observed, with $\Delta^2 \hat p_-=0.50(1)$, or 
$-3.01(9)$~dB, and $\Delta^2 \hat q_+=0.73(1)$, or $-1.37(6)$~dB}
\label{entang}
\end{figure}

Fig.~\ref{entang} shows the recorded noise in sum and subtraction of photocurrent
fluctuations of signal and idler beams as functions of analysis cavities' detuning, 
for $\sigma=1.06$. Off resonance, quantum correlations are observed in the 
subtraction of amplitudes, $\Delta^2 \hat p_-=0.50(1)$, or $-3.01(9)$~dB. For analysis 
cavities' detuning equal to half the bandwidth, squeezing is present in the sum 
of phases, $\Delta^2 \hat q_+=0.73(1)$, or $-1.37(6)$~dB. The Duan {\it et al.} and 
Simon criterion, Eq.(\ref{dgczcrit}), is then clearly violated, 
\begin{equation}
\Delta^2 \hat p_-+\Delta^2 \hat q_+=1.23(2)<2 \;,
\end{equation}
attesting the entanglement. This value, together with the one reported by Jing 
{\it et al.}\cite{pfisterentang}, is the lowest achieved for twin beams produced 
by an above-threshold OPO.

We also point out that, in this experiment, the twin beams have very different 
frequencies (wavelengths differ by $\approx 1$~nm), an unusual situation. Such 
two-color entanglement can be very interesting for the transfer of quantum 
information between different parts of the electromagnetic spectrum. 

\section{Conclusion}

We presented a theoretical and experimental investigation of phase noise and 
entanglement in the above-threshold OPO. Excess noise in the phase sum of the 
twin beams was measured as a function of pump power relative to threshold and 
we found that it decreases as pump power is lowered. We finally discovered that 
excess pump noise is generated inside the OPO cavity containing the nonlinear 
crystal, even for a shot noise limited pump beam and without parametric oscillation. 
The ultimate physical origin of this phenomenon still requires further 
investigation. Another important question to address is how one can eliminate 
this effect. Su {\it et al.}\cite{optlettpeng} were able to observe entanglement 
for $\sigma$ of the order of two. The difference between their setup and others 
is a lower cavity finesse for the pump field. If the assumption of an intensity 
dependent index of refraction is correct, this makes sense. For a lower finesse, 
phase shifts accumulated inside the cavity should be smaller, hence the excess 
noise generated should also be smaller. 

In spite of these unexpected phenomena, two-color entanglement was measured 
in the above-threshold OPO. There are interesting avenues to pursue for applications 
in quantum information. First of all, we should mention that the strongest 
squeezing measured to date, $-9.7$~dB, was generated in an above-threshold 
OPO\cite{claude10db}. Thus, entanglement in the above-threshold OPO may be 
the strongest ever achieved for continuous variables. The bright twin beams 
can have very different frequencies, and one can envisage CV quantum 
teleportation\cite{kimbleteleport} to transfer quantum information from one 
frequency to another (in other words, to ``tune'' quantum information). For 
example, this system could be used to communicate quantum information between 
quantum memories or quantum computers based on ``hardwares'' which have different 
resonance frequencies. Finally, a quantum key distribution protocol proposed 
by Silberhorn {\it et al.}\cite{qkdleuchs} can be readily implemented, with 
the advantage that the measurement with analysis cavities does not require 
sending a local oscillator together with the quantum channel to the distant 
receiver. 

The above-threshold OPO, which was the first system proposed to observe 
continuous variable entanglement, has finally been added to the optical 
quantum information toolbox. We expect new and exciting applications to 
come in the near future. 

\section*{Acknowledgments}

This work was supported by FAPESP, CAPES, and CNPq through {\it Instituto do 
Mil\^enio de Informa\c c\~ao Qu\^antica}. We thank C. Fabre and T. Coudreau for 
kindly lending us the nonlinear crystal.

\end{document}